\documentclass{ptapap}%
\author{A. Pigulski}[IAUWr]
\author{A. Baran}[IFUP]
\author{M. Bzowski}[CBK]
\author{H. Cugier}[IAUWr]
\author{B. Czerny}[CFT]
\author{J. Daszy\'nska-Daszkiewicz}[IAUWr]
\author{W. Dziembowski}[CAMK,OAUW]
\author{G. Handler}[CAMK]
\author{Z. Ko{\l}aczkowski}[IAUWr]
\author{M. Kr\'olikowska}[CBK]
\author{J. Krzesi\'nski}[IFUP]
\author{G. Maciejewski}[CAUMK]
\author{G. Michalska}[IAUWr]
\author{J. Molenda-\.Zakowicz}[IAUWr]
\author{P. Moskalik}[CAMK]
\author{A. Niedzielski}[CAUMK]
\author{E. Niemczura}[IAUWr]
\author{J. Ostrowski}[IAUWr]
\author{A. Pamyatnykh}[CAMK]
\author{M. Ratajczak}[IAUWr]
\author{S. Rucinski}[UoT]
\author{M. Siwak}[IFUP]
\author{R. Smolec}[CAMK]
\author{S. Szutowicz}[CBK]
\author{T. Tomov}[CAUMK]
\author{{\L}. Wyrzykowski}[OAUW]
\author{S. Zo{\l}a}[OAUJ]
\author{M. Sarna}[CAMK]
\affil[IAUWr]{Instytut Astronomiczny, Uniwersytet Wroc{\l}awski, Kopernika 11, 51-622 Wroc{\l}aw, Poland}
\affil[CAMK]{Centrum Astronomiczne im.~M.\,Kopernika PAN, Bartycka 18, 00-716 Warszawa, Poland}
\affil[IFUP]{Instytut Fizyki Uniwersytetu Pedagogicznego, Podchor\k{a}\.zych 2, 30-084 Krak{\'o}w, Poland}
\affil[CBK]{Centrum Bada\'n Kosmicznych PAN, Bartycka 18a, 00-716 Warszawa, Poland}
\affil[CFT]{Centrum Fizyki Teoretycznej PAN, Al.~Lotnik\'ow 32/46, 02-668 Warszawa, Poland}
\affil[OAUW]{Obserwatorium Astronomiczne Uniwersytetu Warszawskiego, Al.~Ujazdowskie 4, 00-478 Warszawa, Poland}
\affil[CAUMK]{Centrum Astronomii, Wydzia{\l} Fizyki, Astronomii i Informatyki Stosowanej, Uniwersytet Miko{\l}aja Kopernika, Grudzi\k{a}dzka 5, 87-100 Toru\'n, Poland}
\affil[UoT]{Department of Astronomy \& Astrophysics, University of Toronto, 50 St.~George Street, Toronto, Canada}
\affil[OAUJ]{Obserwatorium Astronomiczne Uniwersytetu Jagiello\'nskiego, Orla 171, 30-244 Krak\'ow, Poland}
\title{UVSat: a concept of an ultraviolet/optical photometric satellite}
\begin{document}
\maketitle
\def\sat{UVSat}
\def\degr{$^{\rm o}$}
\begin{abstract}
Time-series photometry from space in the ultraviolet can be presently done with only a few platforms, none of which is able to provide wide-field long-term high-cadence photometry. We present a concept of {\sat}, a twin space telescope which will be capable to perform this kind of photometry, filling an observational niche. The satellite will host two telescopes, one for observations in the ultraviolet, the other for observations in the optical band. We also briefly show what science can be done with UVSat.

\end{abstract}
\section{Introduction}
The last two decades in astronomy can be labeled as the time of large surveys. With new detectors, robotic telescopes and dedicated space observatories, we experience now the onset of the era of studies aimed at measuring variability properties of celestial objects. For variability studies the distribution of observations in time is crucial because it defines the time scales, which can be probed. The other important factor is the wavelength range. For obvious reasons, ground-based variability surveys (OGLE, (Super)MACHO, EROS, ASAS, (Super)WASP, CRTS, VVV) were confined to optical and/or near-infrared wavelengths. Space-based projects aimed at monitoring variability were also carried out in visual passband(s) (Hipparcos, MOST, CoRoT, Kepler, BRITE-Constellation, Gaia). Observations in ultraviolet from space were usually carried out in one or several passbands and aimed at characterizing the spectral energy distribution of the observed objects. A typical result of such observations was a single measurement or a small number of measurements per object. Only in some rare cases time-series UV observations of variable objects were provided. Examples of this kind of observation by OAO-2, ANS, Swift's UVOT, GALEX, and HST can be found in the literature \citep{1978ApJ...219..947L,1980ApJ...239..919E,2000ApJ...539..379K,2012AJ....144...84S,2015JHEAp...7..111B}.

The ultraviolet (UV) is the spectral range in which hot objects emit most of their energy. This spectral region is therefore crucial for understanding young and massive objects like OB stars, Wolf-Rayet stars or Luminous Blue Variables, but also for less massive objects at the final stages of their stellar evolution (white dwarfs, hot subdwarfs). Many non-stellar objects, like AGNs, including quasars and blazars, also emit a significant fraction of their energy in the UV. In some circumstances, $\gamma$-ray burst afterglows can be also observed in the UV \citep[e.g.][]{2009ApJ...690..163R}. 

The present-day possibilities of carrying out UV observations are confined to several UV space missions or dedicated experiments on-board X-ray/$\gamma$-ray satellites. These are: Swift with its 30-cm UVOT telescope, Optical Monitor aboard XMM-Newton, and the ASTROSAT mission carrying the twin 37.5-cm UVIT telescopes. Several instruments on-board HST (WFC3, ACS, COS, and STIS) have also the capability of doing UV spectroscopy and photometry. The proposed {\sat} mission (see Sect.~\ref{uvsat}) will have the opportunity to monitor photometrically a few hundred objects for a period of several months. None of the UV missions working or planned for the next decade has a similar ability. With this satellite, we will therefore fill an important observational niche in the study of the variability of hot objects. It can be also complementary to the other large planned UV missions like UVMag/Arago \citep{2015IAUS..307..389N} and WSO-UV \citep{2016ARep...60....1B}. It is worth noting that observations with {\sat} would be carried out at the time when all Gaia results will be available. Knowing distances to stellar objects which will be studied with {\sat}, will bring some of these studies (e.g.~stellar modeling) to a much higher level.

\section{Technical details of {\sat}}\label{uvsat}
From the technical point of view the basic concept of the {\sat} mission can be defined as follows:
\begin{itemize}
\item The satellite will host two refracting telescopes with apertures of $\sim$10~cm, one with optics designed for wide-band UV observations (200\,--\,300~nm), the other, for observations in the visual band (500\,--\,600~nm).
\item The field of view of both telescopes will be the same, and of the order of 10{\degr}\,$\times$\,10{\degr}.
\item The observations will be carried out in selected fields and will last one to six months. In each field, a sample of a few hundred objects will be monitored with a cadence of the order of at least a few seconds. Observing multiple fields during a single orbit is also an option.
\item A large dynamical range will be secured by the ability of on-board stacking or using a detector which allows for an independent red-out of each pixel (sCMOS, CID).
\item The assumed precision of the measurements (per orbit) would be of the order of 1~mmag for $m(\mbox{UV})\approx$~11 mag and $m(\mbox{visual})\approx$~12 mag. Figure 1 shows objects brighter than 9th~mag in 236.5~nm band detected by TD-1 satellite, which defines the bright fraction of objects accessible by UVSat. The number of objects within the reach of UVSat will be, however, much larger and will also include fainter objects, particularly outside the Galactic plane.
\end{itemize}
\begin{figure}[!ht]
\centering
\includegraphics[width=\textwidth]{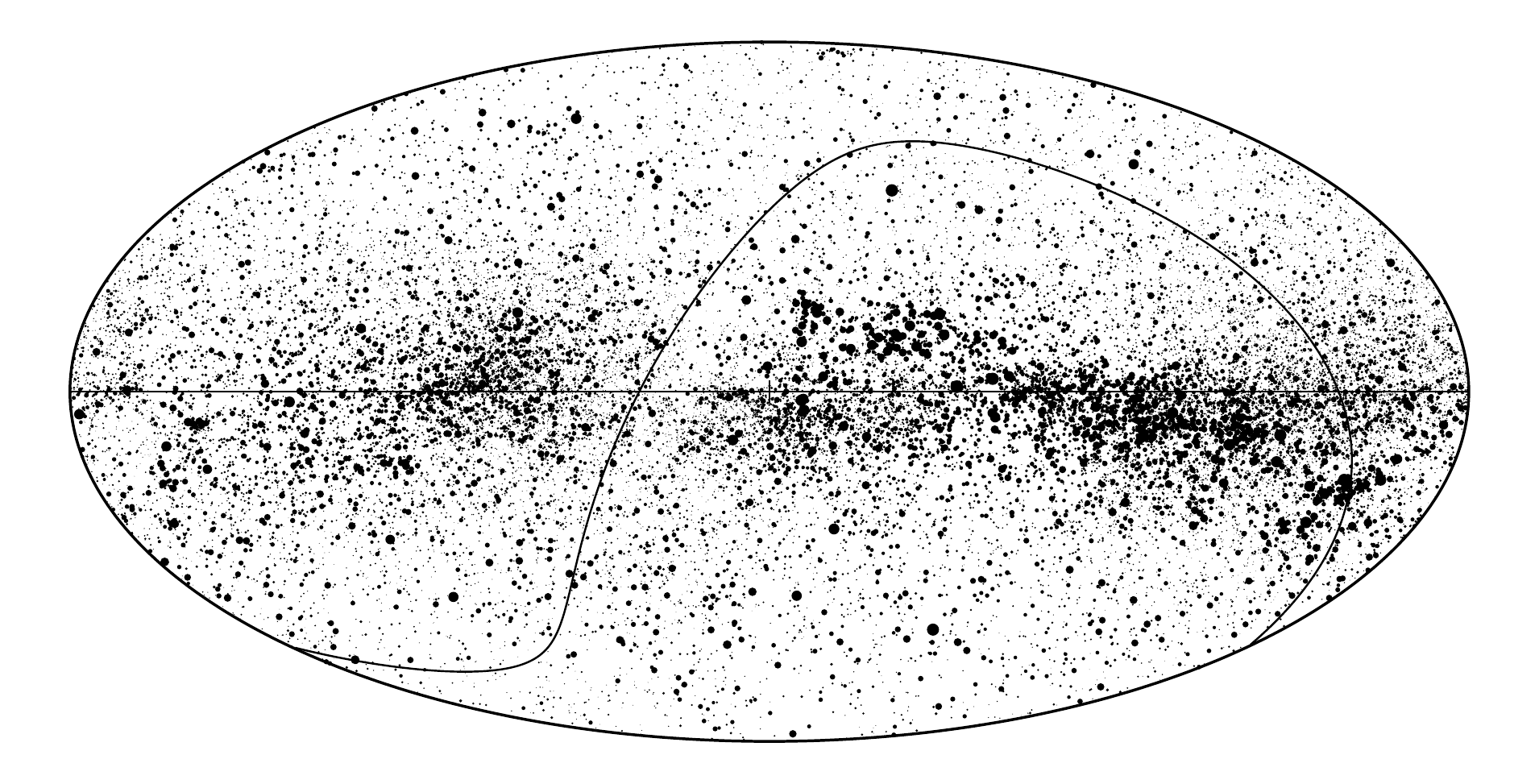}
\caption{Map of the sky in Galactic coordinates (Aitoff projection) containing around 26\,600 objects brighter than $m_\lambda=$~9~mag in the 236.5~nm band of the TD-1 satellite \citep{1978csuf.book.....T}. The horizontal line is the Galactic plane, the curved one, the celestial equator.}
\label{Fig1}
\end{figure}

\begin{figure}[!ht]
\centering
\includegraphics[angle=270,width=0.8\textwidth]{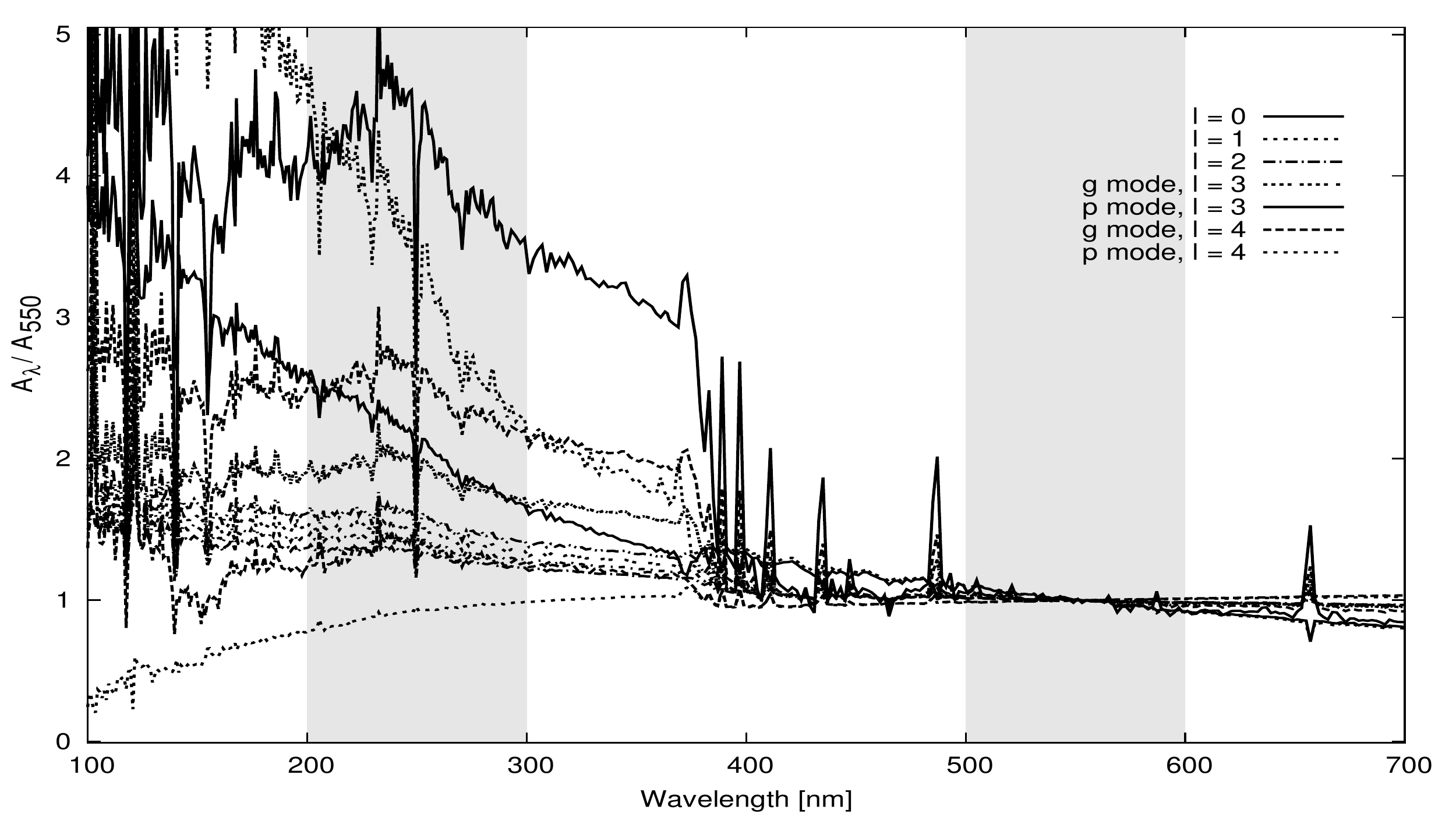}
\caption{\small Monochromatic amplitude ratios (with respect to the amplitude at $\lambda=$\,550\,nm) in the range between 100 and 700~nm for modes with $l=$ 0\,--\,4 and a sample model with stellar mass equal to 10.7\,M$_\odot$. Vertical strips confine the two proposed photometric bands of {\sat}.}
\label{Fig2}
\end{figure}

The possibility of obtaining two-band observations is very important. A similar ability is now enjoyed by BRITE-Constellation \citep[B-C,][]{2014PASP..126..573W}, in which, out of five working satellites two are equipped with a blue ($\lambda_{\rm cen} =$ 425\,nm) filter, while the other three have a red ($\lambda_{\rm cen}=$ 621\,nm) filter. In contrast to B-C, however, {\sat} would be an autonomous observatory. The advantage of observing in two bands, including one in the UV, is that many astrophysical phenomena, which lead to photometric variability, have very different behaviour and/or amplitude in the UV and visual. For example, in hot pulsating stars the variability is mainly due to differences in temperature, which lead to higher amplitudes at shorter wavelengths. This is shown in Fig.~\ref{Fig2}. It is clear from this figure that UV and visual observations will allow for unambiguous identification of the observed modes in terms of the spherical harmonic degree $l$.

\section{Science with {\sat}}
The observations of {\sat} will be focused on UV bright sources; the visual telescope will provide mainly complementary observations. This means that the most interesting targets will be hot objects, predominantly stars. The most important processes related to the chemical evolution of matter in galaxies take place in massive stars. These processes are still not well understood. One of the most attractive possibilities is to use asteroseismology of massive stars to probe their interiors. Asteroseismology can also be used to test stellar physics, e.g.~opacities. In addition, UV time-series photometry can be used to study variability in hot remnants of late stellar evolution, like pre-white dwarfs or hot subdwarfs, and in a variety of interacting and non-interacting binaries with hot components. Finally, some bright extragalactic objects can also be studied in the UV, including AGNs or early stages of supernovae.

A list of the most important objects and phenomena that can be studied with {\sat} is the following:
\begin{itemize}
\item Pulsating hot massive stars. Some distinct problems that can be studied by means of seismic modeling include: (i) the extent of overshooting from the core and its dependence on stellar parameters, (ii) profiles of internal rotation, (iii) finding clues for a review of opacities, which presently do not explain the instability of many observed modes, especially in the g-mode domain, (iv) the role of fast rotation in exciting/damping modes. 
\item The incidence of variability among O-type stars, Wolf-Rayet stars, Luminous Blue Variables, and B-type supergiants.
\item Be stars and the role of pulsations in mass transfer from stellar atmosphere to circumstellar disks (see Baade et al., these proceedings).
\item Massive binaries. Such binaries are more frequent than those consisting of less massive stars \citep{2013ARA&A..51..269D}. Binarity allows for determination of their masses and radii, but also for studies of a diversity of astrophysical phenomena, especially in relatively close systems. Some interesting classes of binary systems like Double Periodic Variables \citep{2010ASPC..435..283M} or high-mass X-ray binaries \citep{2015A&ARv..23....2W} can also be studied.
\item Intermediate and low-mass binaries, including algols and low-mass X-ray binaries.
\item Intermediate-mass main-sequence pulsating stars, $\delta$~Sct and $\gamma$~Dor stars. 
\item Chemically peculiar and rapidly oscillating Ap stars. 
\item Bright symbiotic stars and symbiotic novae. UV time-series observations can be especially useful in the characterization of flickering in these stars originating close to the hot component of a binary \cite[e.g.,][]{1996A&AS..116....1T}.
\item Bright cataclysmic variables (CVs) and novae. These objects will be within the reach of {\sat} only during outbursts. Since time-series UV photometry of CVs is almost non-existent, {\sat} observations would open a new window in observing these stars.
\item Early phases of supernova explosions. In particular, the shock-breakout flares \citep{2010ApJ...725..904N}, important for understanding the characteristics of massive exploding stars, can be observed in the UV. Up-to-date observations are very rare and usually serendipitous \citep{2008Sci...321..223S,2015ApJ...804...28G}.
\item UV bright transients, in particular superluminous supernovae \citep{2011Natur.474..487Q} and tidal disruption events \citep[e.g.,][]{2017MNRAS.465L.114W}.
\item Bright classical pulsating stars (Cepheids, RR Lyrae stars). 
\item Pulsating compact stars (white dwarfs, pre-white dwarfs, and hot subdwarfs).
\item Pre-main sequence stars (classical T Tauri stars, Herbig Ae/Be stars, and FU~Ori stars).
\item Active galactic nuclei (AGNs). Photometric variability of AGNs is very common; lags between light curves in different passbands can help to reveal mechanisms responsible for the variability \cite[e.g.,][]{2014ApJ...788...48S,2015ApJ...806..129E}.
\item Detection of extrasolar planets via the transit method.
\item Occultations by Kuiper Belt objects.
\item Photometric variability of bright comets, including determination of their rotation periods.
\end{itemize}

\acknowledgements{Support from the NCN grants 2015/18/A/ST9/00578 (for G.H.), 2012/05/E/ST9/03915 (for M.S.), 2014/13/B/ST9/00902 (for E.N.~and J.M.-\.Z.), 2015/ 16/S/ST9/00461 (for M.R.), and 2016/21/B/ST9/01126 (for A.P.) is acknowledged. The scientific rationale for {\sat} was prepared as a part of a preliminary feasibility study funded by the Polish Space Agency (POLSA).}

\end{document}